\documentclass[iop]{emulateapj}

\def\lapp{\ifmmode\stackrel{<}{_{\sim}}\else$\stackrel{<}{_{\sim}}$\fi}
\def\gapp{\ifmmode\stackrel{>}{_{\sim}}\else$\stackrel{>}{_{\sim}}$\fi}

\newcommand{\source}{Swift~J1822.3$-$1606}
\newcommand{\src}{Swift~J1822.3$-$1606}
\newcommand{\rxte}{\textit{RXTE}}
\newcommand{\xte}{\textit{RXTE}}
\newcommand{\cxo}{\textit{CXO}}
\newcommand{\chandra}{\textit{Chandra}}
\newcommand{\swift}{\textit{Swift}}
\newcommand{\tempo}{{\tt{TEMPO}}}
\newcommand{\degrees}{^{\circ}}

\newcommand{\myemail}{maggie@physics.mcgill.ca}

\begin{document}

\shorttitle{The Spin-down of Swift J1822.3$-$1606}

\title{The Spin-down of Swift J1822.3$-$1606:  A New Galactic Magnetar}

\author{
M. A. Livingstone\altaffilmark{1},
P. Scholz, V. M. Kaspi, C.-Y. Ng}
\affil{Department of Physics, Rutherford Physics Building,
McGill University, 3600 University Street, Montreal, Quebec,
H3A 2T8, Canada}

\and

\author{Fotis~P.~Gavriil}
\affil{NASA Goddard Space Flight Center, Astrophysics Science
Division, Code
662, Greenbelt, MD 20771, USA}
\affil{Center for Research and Exploration in Space Science and
Technology,
University of Maryland Baltimore County, 1000 Hilltop Circle,
Baltimore, MD 21250, USA}

\altaffiltext{1}{\myemail}

\begin{abstract}
On 2011 July 14, a new magnetar candidate, \src,  was identified via a rate
trigger on the \swift/Burst Alert Telescope. Here we present an
initial analysis of the X-ray properties of the source, using data
from the 
{\textit{Rossi X-ray Timing Explorer}}, \swift,
and the {\textit{Chandra X-ray Observatory}}, spanning 
2011 July 16~--~September 22. 
We measure a precise spin period of $P=8.43771963(5)$\,s and
a spin-down rate of $\dot{P}=2.97(28)\times10^{-13}$,
at MJD~55761.0, corresponding to 
an inferred surface dipole magnetic field strength of $B=5.1 \times 10^{13}$\,G,
the second lowest thus far measured for a magnetar, though similar to
1E~2259+586 as well as to several high-magnetic field radio pulsars. 
We show that the pulsed X-ray flux decay in the 2~--~10\,keV band is best fit by
an exponential with a time constant of 
16.4$\pm$0.3\,days. 
After increasing from $\sim$35\% during
the first week after the onset of the outburst, the pulsed fraction in
the 2~--~10\,keV band 
remained constant at $\sim$45\%. We argue that these properties
confirm this source to be a new member of the class of objects known
as magnetars. 
\end{abstract}


\section{Introduction}

On 2011 July 14, the \swift/Burst Alert Telescope (BAT) triggered on several
bursts of hard X-ray emission from the direction of a previously
unknown source, subsequently named \src\ \citep{cbc+11}.
The spin period of the new source was soon identified to be
$P=8.4377$\,s \citep{gks11}  
and observations with \swift\ X-ray Telescope (XRT) were used to further localize the position of
the source \citep{pbk11}. 

The simplest interpretation of these initial observations is that
\src\ is a magnetar. However, because of suggested similarities between \src\
and the unusual Be X-ray binary J1626$-$5156, namely, a similar pulsed fraction 
and IR counterpart, it was suggested that perhaps \src\
is also a Be X-ray binary \citep{gks11,blm11}. 
This hypothesis was challenged by \citet{hal11}, who, due to the
lack of an optical counterpart \citep{um11},
argued that
the properties of \src\ are in line with those of other magnetars, in
particular, the transient magnetar XTE~J1810$-$197. 
Here we report on timing and flux properties of \src\ that
rule out a Be binary, and instead agree strongly with the magnetar 
identification.

Magnetars are neutron stars powered by the decay of extreme magnetic
fields \citep{td95,td96a,tlk02}.  Soft Gamma-ray
Repeaters (SGRs) and Anomalous X-ray Pulsars (AXPs) are thought to be
observational manifestations of magnetars, and we refer to them as
such here. Magnetars exhibit X-ray bursts,  spin periods ($P$)
in the relatively narrow range\footnote{See the Magnetar Catalog: 
http://www.physics.mcgill.ca/$\sim$pulsar/magnetar/main.html} of
2~--~12\,s, and are observed to spin down with time.  As inferred from $P$ 
and spin-down rate $\dot{P}$ via the conventional vacuum dipole
braking model that implies dipolar magnetic field strength
$B = 3.2 \times 10^{19} (P \dot{P})^{1/2}$\,G,
their inferred dipole magnetic 
fields tend to be significantly larger than for
classical pulsars, typically $B>10^{14}$\,G.  Magnetars are
also observed to display variability in nearly all of their observed properties: 
flux and pulsed flux variations, spectral variations, pulse profile
variability, timing noise and glitches \citep[e.g., see][ for reviews]{mer08,re11}.

The true underlying distribution of magnetic field strengths of
magnetars is an open question. The recent discovery of a magnetar 
with a very low inferred magnetic field of
$B<7.5\times 10^{12}$\,G \citep[SGR~0418+5729; ][]{ret+10},  
has raised the question of how low the dipolar field can be for 
magnetar-like activity to be observed. Moreover, the discovery of
a magnetar-like outburst from the high-$B$ rotation-powered pulsar (RPP)
J1846$-$0258 \citep{ggg+08,ks08a} highlights the question
of whether rotation-powered pulsars and magnetars 
represent distinct classes of object, or if magnetars represent the high-$B$
tail of a single population. 

In this Letter, we present analyses of \textit{Rossi X-ray Timing
Explorer} (\rxte), \swift, and {\textit{Chandra X-ray
Observatory}} (\cxo) data. We perform a phase-coherent timing analysis 
and show that the spin evolution is consistent
with a constant spin-down rate. We present an analysis of the decay of
the source's pulsed intensity after the outburst, and show that the pulsed
fraction is steady after increasing during the first week after the
outburst. We argue that
these properties confirm that \src\ is a new Galactic magnetar. 

\section{Observations}
\label{sec:obs}
\subsection{\swift\ Observations}
On 2011 July 14 at 12:47:47 UT the {\em Swift}/BAT triggered on a rate
increase from the previously unknown source, \source\ \citep{cbc+11}. 
The XRT began observing \source\ on 2011 July 15 at 17:43:20 UT.
A total of 24 observations were taken with the XRT (target IDs 32033
and 32051). The observations span 67\,days: 2011 July
15~--~September 20 (MJD 55758~--~55824), for a total exposure time of
46\,ks.

Cleaned {\em Swift} data in both windowed-timing (WT) mode with
1.8-ms time resolution and photon-counting (PC) mode with 2.5-s time
resolution were downloaded from the {\em Swift} quick-look data
archive\footnote{http://swift.gsfc.nasa.gov/cgi-bin/sdc/ql}.  For 
the lone PC mode observation,
an annular source region with an outer radius of 20 pixels
and an inner radius of 5 pixels was extracted. The inner radius was
excluded because of pile-up.
For the
WT mode observations, 
a 16 pixel strip centered on the source was used. 
Events in the 2~--~10\,keV energy range were then extracted
from the reduced WT mode data
for the timing analysis. The PC mode data have
insufficient time resolution to be useful for timing. 

\subsection{\rxte\ Observations}
\label{sec:rxteobs}

\src\ data were obtained with the Proportional
Counter Array \citep[PCA;][]{jmr+06} on board \rxte. The PCA consists of five
xenon/methane proportional counter units (PCUs) 
sensitive to 2~--~60\,keV photons, with an effective area of
$\sim \rm{6500\,cm^2}$ and a  $\rm{\sim 1^o}$\,FWHM field of view. 
We downloaded 23 public \rxte\ observations of \src\
from the HEASARC archive\footnote{http://heasarc.gsfc.nasa.gov/docs/archive.html} 
(observing program P96048). Data 
were collected in  {\tt{GoodXenon}} mode, which records the arrival time
(1-$\mu$s resolution) and energy (256-channel resolution) of each
event. The observations span 62\,days: 2011 July 16~--~September 16 
(MJD~55758~--~55820), for a total exposure time of 119\,ks.

For our timing analysis, we extracted 2~--~10\,keV photons 
(because this produced high significance pulse profiles
for individuals observations) from the top xenon layer of each PCU
to reduce contamination from the particle background. Data
from individual PCUs were then merged.
When more than one observation
occurred in a 24-hr period, we combined the data to 
produce better profiles. 

\subsection{ {\textit{Chandra}} Observations}
\label{sec:cxoobs}

\cxo\ ACIS-S observations were made on 2011 July 27 (MJD
55769.2,  ObsID 12612), August 4 (MJD 55777.1,  ObsID 12613),  
and September 18 (MJD 55822.7, ObsID 12614)
with net exposures of 
15.0\,ks, 13.7\,ks, and 10.0\,ks, respectively. All three observations were
obtained in continuous clocking  mode, with
 2.85-ms time resolution. 
We carried out the data reduction using CIAO 4.3 with CALDB
4.4.3\footnote{http://cxc.harvard.edu/ciao4.3}. 
We reprocessed the data to retain events on the chip node boundary, then
extracted the source counts from a 6\arcsec-width aperture and the
0.5--8\,keV energy range for our timing analysis, and 2--8\,keV for
the pulsed fraction analysis. 
 
\section{Analysis \& Results}

\subsection{Phase-coherent Timing Analysis}
\label{sec:timing}

Our timing analysis of \src\ follows the common phase-coherent approach,
in which we account for each rotation of the pulsar \citep[see, for
example][]{mt77} and utilized a combination of timing data from \rxte,
\swift, and \chandra. For data from each telescope, events were reduced to barycentric
dynamical time (TDB) at the solar system barycenter using the XRT
position of RA=$18^{\rm{h}}$~$22^{\rm{m}}$\,$18^{\rm{s}}$,
Dec=$-16\degrees$\,$04\arcmin$\,$26.8\arcsec$ \citep[J2000; ][]{pbk11}
and the JPL DE200 solar system ephemeris.
Events were then binned into time series with 31.25-ms time resolution. 

Each \rxte\ (2~--~10\,keV), \swift\ (2~--~10\,keV),  and \cxo\ (0.5~--~8\,keV)
time series was folded with 64 phase bins using a frequency
determined from a periodogram analysis. 
After finding an initial
phase-coherent timing solution, we used this ephemeris to
re-fold all the profiles to produce higher quality 
pulse Times Of Arrival (TOAs).
For \rxte\ data, we found that using 128 phase bins
created good pulse profiles with optimal TOA uncertainties.
For \swift\ and \cxo\ data, 64-bin profiles resulted
in higher significance pulse profiles and good TOA uncertainties.
128- and 64-bin template profiles were created by aligning  and
summing all \rxte\ profiles.

To account for the different energy ranges and
telescope responses between \rxte, \swift, and \cxo, 
which result in small but significant differences in the 
profiles, we fit for a constant phase offset 
between TOAs obtained from different telescopes. 
We used a subset
of the available data to fit for the phase offsets (MJD~55758~--~55781),
which were then held fixed for the remainder of the timing analysis. 

We searched for time variability in the pulse profile and found that
individual profiles are consistent with the template in each case
except for the first observation after the outburst (96048-02-01-00). 
In this case, the variation is subtle, at the $\sim$0.015 phase level. 
We accounted for this by multiplying the corresponding TOA uncertainty by three. 

For each profile, we implemented a
Fourier domain filter by using six harmonics in the cross-correlation
procedure with the appropriate 64- or 128-bin template.
Cross-correlation produces a TOA
for each observation with a typical uncertainty of $\sim$27\,ms
(0.32\% of $P$) for \rxte\
TOAs, $\sim$59\,ms (0.70\% of $P$) for \swift\
TOAs,  and $\sim$23\,ms (0.27\% of $P$) for \cxo\ TOAs.
TOAs were fitted with the software package
{\tt{TEMPO}}\footnote{http://www.atnf.csiro.au/people/pulsar/tempo/}; 
parameters from this fit are given in Table~\ref{table:coherent}. 

\begin{figure}
\plotone{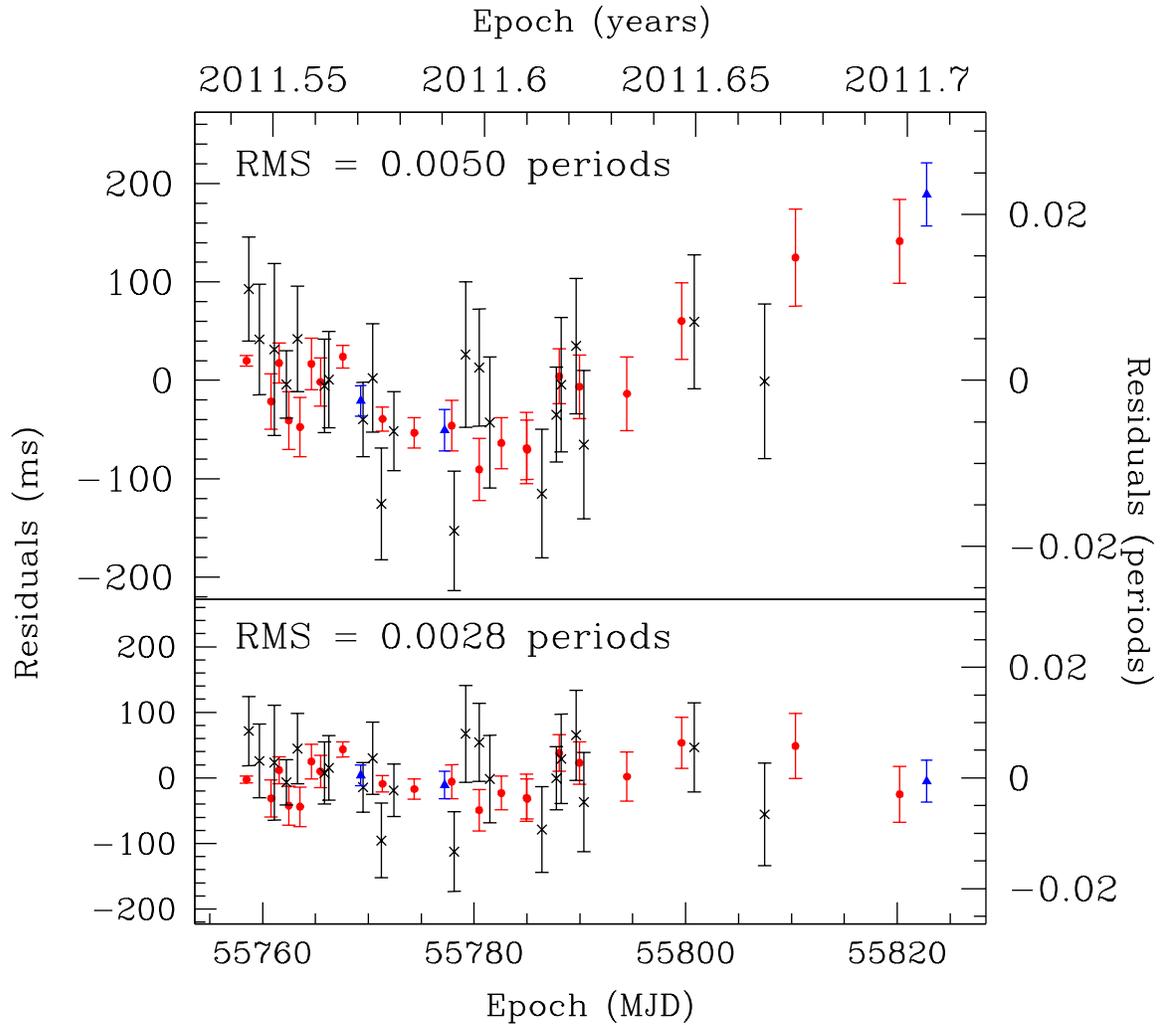}
\figcaption[]{X-ray timing residuals for \src\
spanning 2011 July 16~--~September 22. 
\rxte\ points are shown as filled red circles, \swift\ points are black crosses,
and \cxo\ points are blue triangles. The top panel shows 
residuals with only $\nu$ fitted, while the bottom panel
shows timing residuals with $\dot\nu$ also fitted.
Parameters from this fit are given in Table~\ref{table:coherent}.
\label{fig:resids}}
\end{figure}

The top panel of Figure~\ref{fig:resids} shows timing residuals with
only spin-frequency ($\nu$) fitted, while the bottom panel shows
residuals with the frequency derivative ($\dot\nu$) also fitted. 
Fitting $\dot\nu$ improves the 
RMS residuals by a factor of 1.8 and improves the ${{\chi^2}_\nu}/\nu$
from 3.66/44 to 1.14/43. Resulting phase residuals 
show minimal evidence for an unmodelled trend and fitting further
parameters does not significantly improve the fit. 
The fitted values of $\nu$ and $\dot\nu$
imply a surface dipolar magnetic field of $B=5.1\times 10^{13}$\,G.

\subsection{Pulsed Count-rate Evolution}

To measure the source intensity and decay after the
outburst, we first checked for the
presence of dust scattering rings, such as those observed around
1E~1547$-$5408 \citep{tve+10}, because these will bias the inferred
source
count rate. We checked for deviations from the point-spread 
functions (PSFs) in the first \swift\ PC mode and \cxo\ 
observations following the
outburst. The radial profiles were well fit by the model PSFs, hence we 
found no evidence for dust scattering rings.

To measure the pulsed count rate for each \swift\ and \cxo\
observation, the barycentered time series were folded with the
ephemeris (see Table \ref{table:coherent} and 
Section~\ref{sec:timing}), with 16 phase bins. 
The pulsed count rate for each folded profile in the 2~--~10\,keV energy
range was then determined using a root-mean-squared (RMS) method. 
We calculate the RMS pulsed count rate using Fourier components as
described in \citet{dkg08}, using 5 Fourier harmonics of the pulse
profile.  The pulsed fraction was then
determined by dividing the pulsed count rate by the
total source count rate.

To measure the pulsed count rate for \rxte\ data, 
we created phase-resolved spectra with 16 bins using the software
program {\tt{fasebin}}\footnote{http://heasarc.nasa.gov/docs/xte/recipes/fasebin.html}
to fold barycentered photons with the ephemeris
(Table~\ref{table:coherent}).
We selected photons in the 2~--~10\,keV energy range and from the
top xenon detection layer. To minimize instrumental effects,
we included only data from PCU2.
PCU0 and PCU1 have lost propane layers, adversely affecting their
particle background
levels\footnote{http://heasarc.gsfc.nasa.gov/docs/xte/pca\_history.html},
and data available from PCU3 and PCU4 were minimal for this
source. The RMS method with 5 harmonics was then used to measure the pulsed
count rate. Because the PCA is not a focusing instrument the pulsed
fraction could not be accurately determined from these
data. 

\begin{figure}
\plotone{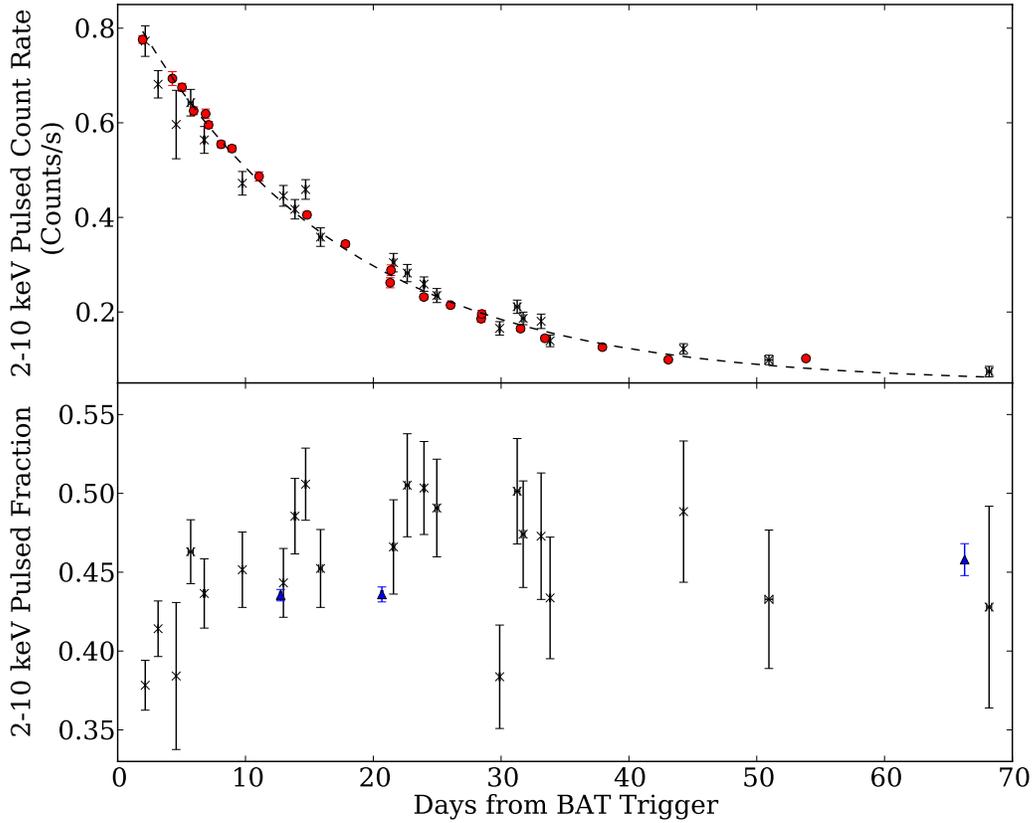}
\figcaption{Top: Pulsed count rate evolution of \src\ in the 2~--~10\,keV energy range
following its outburst. \swift\ data are shown as black crosses,
\xte\ data are filled red circles. The
dashed line indicates the best-fit exponential decay with
$\tau=16.4\pm0.3$\,days. Bottom: Pulsed fraction measurements in the
2~--~10\,keV with \swift\ data as black crosses and \cxo\ data as blue 
triangles. \label{fig:flux} }
\end{figure}

Figure~\ref{fig:flux} shows the evolution of the pulsed source
intensity (top panel, as measured with \swift\ and \rxte), and
the pulsed fraction (bottom panel, as measured with \swift\ and 
\cxo). We present here the
pulsed source intensity evolution only in terms of count rate, instead
of
fluxes. A detailed spectral analysis will follow in a subsequent paper.
To describe this pulsed intensity decay quantitatively, we fit the
pulsed count-rate evolution of \src\ in the 2~--~10\,keV energy range
with an exponential decay. Pulsed count rates from a given source for
\xte\
and \swift\ are different owing to instrumental effects (e.g.
effective area), but should be offset only by a constant, 
and trends should be
the same between instruments (as is observed here).
Thus, \xte\ PCA pulsed count rates were scaled to the \swift\
XRT pulsed count rates. The scaling factor 
was chosen to be that which minimized the residuals
of the fit to the pulsed count rates.
The exponential decay is described by $F(t) = F_p \exp^{-(t-t_0)/\tau}
+ F_q$ where $F_p$
is the peak count rate, $F_q$ is the count rate in quiescence,
$t_0$ is the time of the BAT trigger in MJD and $\tau$ is the decay
timescale in days. With a $\chi^2_\nu$ of 2.69
for 41 degrees of freedom, the best-fit decay timescale is
$16.4\pm 0.3$ days. A power-law model was also fit to the decay,
but it provided a much worse fit with a $\chi^2_\nu$ of 79 for 41
degrees of freedom.

The bottom panel of Figure \ref{fig:flux} shows the evolution of the
2~--~10 keV pulsed fraction of \src\ for \swift\ and \cxo\ data.
The pulsed fraction appears to have increased from $\sim$35\%
immediately after the outburst until
about one week after the BAT trigger. It then remained constant at
$\sim$45\%.

\section{Discussion}
\label{sec:disc}
The regular spin-down we report for \src, together with a pulsed flux
decay similar to that seen in other magnetar outbursts, demonstrates that
this source indeed shares critical properties with other known
magnetars; we hence classify it as such. The inferred surface dipole $B$ for \src\ of $5.1\times
10^{13}$\,G is smaller than those of all but one confirmed
magnetar, SGR~0418+5729 \citep{ret+10}, though it is very close to that of
AXP 1E~2259+586 \citep[$B=5.9\times 10^{13}$\,G;][]{gk02}.
 As for the latter,
\src's $B$ is similar to that measured 
for several RPPs, including the lone 
identified magnetically 
active RPP, PSR~J1846$-$0258 which has $B=4.9 \times 10^{13}$\,G
\citep{gvb+00}. 
Although enhanced spin-down could be present if the source suffered a
large glitch at the outburst \citep[e.g. ][]{kgw+03}, the duration of
our observations of \src\ (64\,days) significantly exceeds the glitch
recovery time scales measured in previous large magnetar glitches, so
any contamination is  likely to be small.

\begin{figure}
\plotone{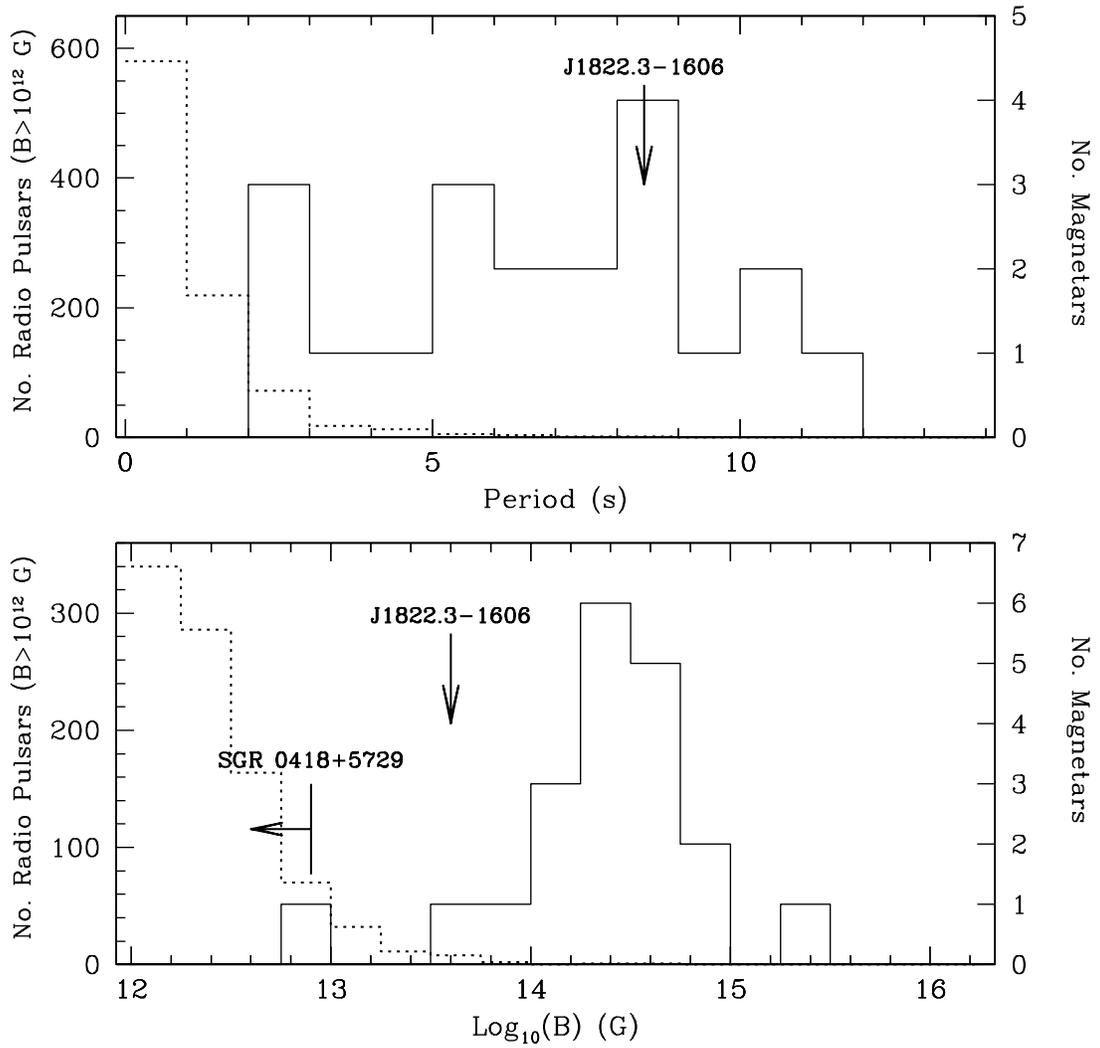}
\figcaption{
(Top) Histograms of the spin periods of observed magnetars (for which $\dot{P}$ has also been measured;
see the online magnetar catalog, http://physics.mcgill.ca/$\sim$magnetar/main.html; solid) and known
radio pulsars having inferred $B>10^{12}$\,G (from the ATNF pulsar catalog at
http://www.atnf.csiro.au/research/pulsar/psrcat/; dotted).  \src\ is indicated.
Note the different vertical axes for the radio pulsars (left) and magnetars (right).
(Bottom) Histograms of the inferred $B$-field strength of the
same magnetars (solid) and the same radio pulsars (dotted).  \src\ is indicated, as is
the upper limit for SGR~0418+5729 \citep{ret+10}.  Note again the different vertical axes.
\label{fig:PBdist}
}
\end{figure}

Interestingly, the most recently discovered magnetars (including
SGR~0418+5729 and 
\src, but also SGRs~0501+4516, 1833$-$0832 and Swift~J1834.9$-$0846;
G{\"o}{\u g}{\"u}{\c s} et al. 2010a,b; Rea et al. 2010, Kuiper \&
Hermsen 2011)\nocite{gwk+10,gcl+10,ret+10,kh11} all have 
$B \lapp 1 \times 10^{14}$\,G, effectively lowering the `average'
inferred $B$-field for magnetars.  This raises the question of
what is the true distribution of magnetar field strengths.
In Figure~\ref{fig:PBdist}, we plot the distributions of the periods and
inferred surface dipolar magnetic fields of all confirmed magnetars, as
well as of all known $B>10^{12}$\,G radio pulsars.  

The $P$ distribution
of magnetars remains narrow, spanning less than an order of magnitude, in great
contrast to those of radio pulsars.  We note that the period distribution for magnetars
is statistically consistent with being flat; a fit to the histogram mean yields a ${\chi^2}_\nu$
of 0.63.  This is somewhat surprising and needs to be
addressed in any future population studies. 
The $B$ distribution for observed magnetars, by contrast,
appears more peaked, even with the relatively low $B$ we have measured for \src,
albeit with the possible low tail represented by SGR~0418+5729.  
If the magnetars
represent the high-$B$ tail of the $B$ distribution of all non-recycled neutron
stars, then their peaked $B$ distribution may indicate that 
burst rates of these sources fall rapidly with decreasing $B$, since the primary
mechanism for their discovery is through their bursting behavior,
particularly with the \swift\ BAT and {\textit{Fermi}} GBM.  This would be
broadly consistent with the theoretical findings of \citet{pp11}, who calculate expected
burst rates in neutron stars and show that
all neutron stars can show magnetar-like bursts, though the burst rate drops with
decreasing $B$ and increasing age.
That the distribution of radio pulsars plummets rapidly above
$B\sim 10^{13}$\,G
could genuinely be due to fall-off in the intrinsic distribution, or
to smaller radio 
beams for these preferentially longer-period pulsars, but may also
indicate that radio emission is harder to produce at higher $B$ \citep[e.g.][]{bh01}.  
Although radio emission has been observed for three magnetars
\citep{crh+06,crhr07,lbb+10}, their radio spectra and radiative 
stabilities are vastly different than for conventional radio pulsars, 
suggesting a distinct emission mechanism.  Hence, if indeed we can `unify' radio
pulsars and magnetars as described above, we expect the lowest $B$
magnetars to be the likeliest
to produce observable traditional radio-pulsar-like radio emission, and the
highest $B$ RPPs to be the likeliest to show magnetar-like X-ray outbursts.  
The latter is consistent
with the behavior of PSR J1846$-$0258 \citep{ggg+08}, while as yet,
there is no evidence to support the former prediction.
Such a unification would also suggest that
there exists a large population of as yet undetected $B\sim 10^{13}$\,G
neutron stars that could be one day detected via their occasional
magnetar-like bursts.

The characteristic age ($\tau_c=P/2\dot{P}$) for \src\ is 450\,kyr, comparable to that
of AXP 1E~2259+586  but much larger than for all other magnetars, except for
SGR~0418+5729 ($>$24\,Myr).  1E~2259+586 is firmly associated with the supernova remnant CTB 109
which is thought to have an age of $\sim$9\,kyr \citep{spg+04}, 
much smaller than the AXP's characteristic age of 250\,kyr. 
This demonstrates that $\tau_c$ for magnetars can be 
unreliable true age indicators, which perhaps can be ascribed to epochs of greater spin-down torque early in
the star's history, when the dipolar field was larger.  Indeed this hypothesis
can also explain the large $\tau_c$ of SGR~0418+5729, although the latter's
greater distance from the Galactic Plane suggests it may indeed be the oldest
known magnetar \citep[but see][for an alternative discussion]{aec11}.
\citet{tzp+11} suggest that the observed properties of SGR~0418+5729
are consistent with it being an aged magnetar in which the external $B$ has decayed
significantly. Further evidence
for SGR~0418+5729 being older than other magnetars includes lower energy
bursts and a very low quiescent luminosity. Whether
\src\ more closely resembles SGR~0418+5729 or 1E~2259+586 remains to
be seen. The measurement of its flux in quiescence and the  
discovery of an associated supernova remnant would
help resolve this issue.

The post-outburst pulsed count rate evolution of \src\ is best characterized by an
exponential decay with timescale of 16.4$\pm$0.3\,days. This is
similar to what has been observed after several other magnetar
outbursts \citep[e.g.][]{rit+09,gh05,ggg+08}, though power-law
decays are also common \citep[e.g.][]{wkg+01,kew+03}. 
\citet{let02} predicted power-law decays in magnetar
outbursts assuming crustal cooling following an impulsive heat injection.  
\citet{bt07}, by contrast, considered magnetar outbursts as a result of sudden twisting of
magnetic field lines in the magnetosphere, with the relaxation a result of their
untwisting.  In this model,
the relaxation is predicted to be approximately linear, a functional
form observed in one outburst of 1E~1048.1$-$5937 \citep{dkg09}.
\citet{bel09} showed that other functional forms for the decay, including exponential, are
possible, as the untwisting
is predicted to be strongly non-uniform, being erased by a propagating
`front' whose speed
depends on the initial twist configuration.  The diversity in
functional forms for the flux decays of 
magnetars hence favors the model put forth by \citet{bel09}.

\acknowledgments
We thank the {\textit {Chandra}} and \swift\ teams for their work in
scheduling Target of Opportunity observations.
This research made use of data obtained from the High Energy
Astrophysics Science Archive Research Center Online Service, provided by 
NASA-GSFC. V.M.K. holds the Lorne Trottier Chair in Astrophysics and
Cosmology and a Canada Research Chair in Observational Astrophysics. 
This work is supported by NSERC via 
a Discovery Grant, by FQRNT, by CIFAR, and a Killam Research Fellowship.
CYN is a Tomlinson postdoctoral fellow and a CRAQ postdoctoral fellow.

\normalsize

\begin{deluxetable}{lc}
\tablecaption{Spin Parameters for \src.
\label{table:coherent}}
\tablewidth{0pt}
\tablehead{\colhead{Parameter} & \colhead{Value}}
\startdata
Dates (Modified Julian Day)         & 55758~--~55822\\
Epoch (Modified Julian Day)         & 55761.0 \\
Number of TOAs - \rxte\              & 21\\
Number of TOAs - \swift\             & 22 \\
Number of TOAs - \cxo\               & 3 \\
$\nu$ (Hz)                          & 0.1185154336(8)\\
$\dot{\nu}$ (Hz$^2$)                & $-4.2(4)\times10^{-15}$ \\
RMS residuals (ms)                  & 23.4 \\
RMS residuals (periods)             & 0.0028 \\
\cutinhead{Derived Parameters}  
$P$ (s)                                          & 8.43771963(5)  \\
$\dot{P}$                                        & $2.97(28)\times10^{-13}$ \\
Surface Dipolar Magnetic Field, $B$ (G)          & $5.1(2)\times10^{13}$ \\
Spin-down Luminosity, $\dot{E}$ (erg\,s$^{-1}$)  & $2.0(2)\times10^{31}$ \\
Characteristic age, $\tau_c$ (kyr)               & 450(40)     \\
\enddata
\tablecomments{Uncertainties are formal 1$\sigma$ \tempo\ errors.}
\end{deluxetable}


\begin{thebibliography}{40}
\expandafter\ifx\csname natexlab\endcsname\relax\def\natexlab#1{#1}\fi

\bibitem[{{Alpar} {et~al.}(2011){Alpar}, {Ertan}, \& {{\c C}al{\i}{\c  s}kan}}]{aec11}
  {Alpar}, M.~A., {Ertan}, {\"U}., \& {{\c C}al{\i}{\c s}kan}, {\c S}.
  2011, \apjl, 732, L4
    
 \bibitem[{{Bandyopadhyay} {et~al.}(2011){Bandyopadhyay}, {Lucas}, \&
   {Maccarone}}]{blm11}
 {Bandyopadhyay}, R.~M., {Lucas}, P.~W., \& {Maccarone}, T. 2011,
  The Astronomer's Telegram, 3502
	
\bibitem[{{Baring} \& {Harding}(2001)}]{bh01}
{Baring}, M.~G. \& {Harding}, A.~K. 2001, ApJ, 547, 929
	
\bibitem[{{Beloborodov}(2009)}]{bel09}
{Beloborodov}, A.~M. 2009, \apj, 703, 1044
	
\bibitem[{{Beloborodov} \& {Thompson}(2007)}]{bt07}
{Beloborodov}, A.~M. \& {Thompson}, C. 2007, ApJ, 657, 967
	
\bibitem[{{Camilo} {et~al.}(2006){Camilo}, {Ransom},
{Halpern}, {Reynolds}, {Helfand}, {Zimmerman}, \& {Sarkissian}}]{crh+06}
 {Camilo}, F., {Ransom}, S., {Halpern}, J., {Reynolds}, J.,
 {Helfand}, D., {Zimmerman}, N., \& {Sarkissian}, J. 2006, Nature, 442, 892
	    
 \bibitem[{{Camilo} {et~al.}(2007){Camilo}, {Ransom},
   {Halpern}, \&  {Reynolds}}]{crhr07}
   {Camilo}, F., {Ransom}, S.~M., {Halpern}, J.~P., \&
   {Reynolds}, J. 2007, ApJ,  666, L93
		
\bibitem[{{Cummings} {et~al.}(2011){Cummings},
{Burrows}, {Campana}, {Kennea},
  A., , {Palmer}, T., \& {Zane}}]{cbc+11}
  {Cummings}, J.~R., {Burrows}, D., {Campana}, S.,
  {Kennea}, J.~A., A., K.~H., ,
   {Palmer}, D.~M., T., S., \& {Zane}, M. 2011, GRB
    Circular Network, 12159
		    
\bibitem[{{de Ugarte Postigo} \& {Munoz-Darias}(2011)}]{um11}
   {de Ugarte Postigo}, A. \& {Munoz-Darias}, T.
   2011, The Astronomer's Telegram,   3518, 1

\bibitem[{{Dib} {et~al.}(2008){Dib}, {Kaspi}, \& {Gavriil}}]{dkg08}
{Dib}, R., {Kaspi}, V.~M., \& {Gavriil}, F.~P. 2008, ApJ, 673, 1044

\bibitem[{{Dib} {et~al.}(2009){Dib}, {Kaspi}, \& {Gavriil}}]{dkg09}
---. 2009, \apj, 702, 614

\bibitem[{{Gavriil} {et~al.}(2008){Gavriil}, {Gonzalez}, {Gotthelf},
{Kaspi},
  {Livingstone}, \& {Woods}}]{ggg+08}
  {Gavriil}, F.~P., {Gonzalez}, M.~E., {Gotthelf}, E.~V., {Kaspi},
  V.~M.,
    {Livingstone}, M.~A., \& {Woods}, P.~M. 2008, Science, 319, 1802
    
\bibitem[{{Gavriil} \& {Kaspi}(2002)}]{gk02}
{Gavriil}, F.~P. \& {Kaspi}, V.~M. 2002, ApJ, 567, 1067
    
\bibitem[{{Gotthelf} \& {Halpern}(2005)}]{gh05}
{Gotthelf}, E.~V. \& {Halpern}, J.~P. 2005, \apj, 632, 1075
    
\bibitem[{{Gotthelf} {et~al.}(2000){Gotthelf}, {Vasisht},
  {Boylan-Kolchin}, \& {Torii}}]{gvb+00}
  {Gotthelf}, E.~V., {Vasisht}, G., {Boylan-Kolchin}, M., \&
 {Torii},  K. 2000,
	      \apjl, 542, L37
	      
 \bibitem[{{G{\"o}{\u g}{\"u}{\c s}}
 {et~al.}(2010{\natexlab{a}}){G{\"o}{\u  g}{\"u}{\c s}},
 {Cusumano}, {Levan}, {Kouveliotou}, {Sakamoto}, {Barthelmy},
 {Campana}, {Kaneko}, {Stappers}, {de Ugarte Postigo},
 {Strohmayer},{Palmer}, {Gelbord}, {Burrows}, {van der Horst},
 {Mu{\~n}oz-Darias}, {Gehrels},
 {Hessels}, {Kamble}, {Wachter}, {Wiersema}, {Wijers},
 \&{Woods}}]{gcl+10}
 {G{\"o}{\u g}{\"u}{\c s}}, E., et al.
 2010{\natexlab{a}}, \apj, 718,  331
		  
 \bibitem[{{G{\"o}{\u g}{\"u}{\c s}}
  {et~al.}(2011){G{\"o}{\u g}{\"u}{\c s}},
   {Kouveliotou}, \& {Strohmayer}}]{gks11}
   {G{\"o}{\u g}{\"u}{\c s}}, E., {Kouveliotou}, C.,
   \& {Strohmayer}, T. 2011, The Astronomer's Telegram, 3491
		      
\bibitem[{{G{\"o}{\u g}{\"u}{\c s}}
{et~al.}(2010{\natexlab{b}}){G{\"o}{\u  g}{\"u}{\c s}}, {Woods}, 
{Kouveliotou}, {Kaneko}, {Gaensler}, \& {Chatterjee}}]{gwk+10}
 {G{\"o}{\u g}{\"u}{\c s}}, E., {Woods}, P.~M., {Kouveliotou}, C.,
{Kaneko}, Y.,  {Gaensler}, B.~M., \& {Chatterjee}, S. 2010{\natexlab{b}}, \apj,
  722, 899
      
\bibitem[{{Halpern}(2011)}]{hal11}
 {Halpern}, J. 2011, GRB Coordinates Network, 12170, 1
      
\bibitem[{{Jahoda} {et~al.}(2006){Jahoda}, {Markwardt},
 {Radeva}, {Rots}, {Stark}, {Swank}, {Strohmayer}, \& {Zhang}}]{jmr+06}
 {Jahoda}, K., {Markwardt}, C.~B., {Radeva}, Y., {Rots}, A.~H.,
{Stark}, M.~J.,  {Swank}, J.~H., {Strohmayer}, T.~E., \& {Zhang}, W. 2006,
  ApJS, 163, 401
	  
\bibitem[{Kaspi {et~al.}(2003)Kaspi, Gavriil, Woods, Jensen,
  Roberts, \&  Chakrabarty}]{kgw+03}
    Kaspi, V.~M., Gavriil, F.~P., Woods, P.~M., Jensen, J.~B.,
    Roberts, M. S.~E., \& Chakrabarty, D. 2003, ApJ, 588, L93
	      
\bibitem[{{Kouveliotou} {et~al.}(2003){Kouveliotou},
 {Eichler}, {Woods},{Lyubarsky}, {Patel}, {G{\" o}{\u g}{\" u}{\c s}},
{van der Klis}, {Tennant},   {Wachter}, \& {Hurley}}]{kew+03}
 {Kouveliotou}, C., {Eichler}, D., {Woods}, P.~M.,
 {Lyubarsky}, Y., {Patel},
 S.~K., {G{\" o}{\u g}{\" u}{\c s}}, E., {van der
 Klis}, M., {Tennant}, A., {Wachter}, S., \& {Hurley}, K. 2003, ApJ, 596, L79
		      
\bibitem[{{Kuiper} \& {Hermsen}(2011)}]{kh11}
{Kuiper}, L. \& {Hermsen}, W. 2011, The Astronomer's Telegram, 3577, 1
		      
\bibitem[{{Kumar} \& {Safi-Harb}(2008)}]{ks08a}
{Kumar}, H.~S. \& {Safi-Harb}, S. 2008, ApJL, 678, L43
		      
\bibitem[{{Levin} {et~al.}(2010){Levin},
{Bailes}, {Bates}, {Bhat}, {Burgay}, {Burke-Spolaor}, {D'Amico}, {Johnston},
 {Keith}, {Kramer}, {Milia}, {Possenti}, {Rea}, {Stappers}, \& {van Straten}}]{lbb+10}
 {Levin}, L., et al. 2010, \apjl, 721, L33
			   
\bibitem[{{Lyubarsky}  {et~al.}(2002){Lyubarsky}, {Eichler}, \&
    {Thompson}}]{let02}
{Lyubarsky}, Y., {Eichler}, D., \& {Thompson}, C. 2002, ApJ, 580, L69
    
\bibitem[{Manchester \& Taylor(1977)}]{mt77}
Manchester, R.~N. \& Taylor, J.~H. 1977, Pulsars (San Francisco:
Freeman)

\bibitem[{{Mereghetti}(2008)}]{mer08}
{Mereghetti}, S. 2008, \aapr, 15, 225

\bibitem[{{Pagani} {et~al.}(2011){Pagani}, {Beardmore}, \&
{Kennea}}]{pbk11}
{Pagani}, C., {Beardmore}, A.~P., \& {Kennea}, J.~A. 2011, The
Astronomer's
  Telegram, 3493
  
\bibitem[{{Perna} \& {Pons}(2011)}]{pp11}
{Perna}, R. \& {Pons}, J.~A. 2011, \apjl, 727, L51
  
\bibitem[{{Rea} \& {Esposito}(2011)}]{re11}
{Rea}, N. \& {Esposito}, P. 2011, in High-Energy Emission from
Pulsars and their Systems, ed. {D.~F.~Torres \& N.~Rea}, 247
    
    
\bibitem[{{Rea} {et~al.}(2010){Rea}, {Esposito}, {Turolla},
{Israel}, {Zane},  {Stella}, {Mereghetti}, {Tiengo}, {G{\"o}tz}, 
{G{\"o}{\u g}{\"u}{\c s}}, \&   {Kouveliotou}}]{ret+10}
{Rea}, N., et al. 2010, Science, 330, 944
      
\bibitem[{{Rea} {et~al.}(2009){Rea}, {Israel}, {Turolla},{Esposito},
  {Mereghetti}, {G{\"o}tz}, {Zane}, {Tiengo}, {Hurley}, {Feroci},
{Still}, {Yershov}, {Winkler}, {Perna}, {Bernardini},
 {Ubertini}, {Stella},  {Campana}, {van der Klis}, \& {Woods}}]{rit+09}
  {Rea}, N., et al. 2009, \mnras, 396, 2419
	  
\bibitem[{{Sasaki} {et~al.}(2004){Sasaki}, {Plucinsky},
 {Gaetz}, {Smith}, {Edgar}, \& {Slane}}]{spg+04}
   {Sasaki}, M., {Plucinsky}, P.~P., {Gaetz}, T.~J., {Smith},
   R.~K., {Edgar}, R.~J., \& {Slane}, P.~O. 2004, \apj, 617, 322
	      
\bibitem[{{Thompson} \& {Duncan}(1995)}]{td95}
{Thompson}, C. \& {Duncan}, R.~C. 1995, MNRAS, 275, 255
	      
\bibitem[{Thompson \& Duncan(1996)}]{td96a}
 Thompson, C. \& Duncan, R.~C. 1996, ApJ, 473, 322

\bibitem[{Thompson {et~al.}(2002)Thompson, Lyutikov, \&
Kulkarni}]{tlk02}
Thompson, C., Lyutikov, M., \& Kulkarni, S.~R. 2002, ApJ, 574, 332

\bibitem[{{Tiengo} {et~al.}(2010){Tiengo}, {Vianello}, {Esposito},
{Mereghetti}, {Giuliani}, {Costantini}, {Israel}, {Stella},{Turolla},
 {Zane}, {Rea}, {G{\"o}tz}, {Bernardini}, {Moretti}, {Romano},
 {Ehle}, \& {Gehrels}}]{tve+10}
{Tiengo}, A., et al. 2010, ApJ, 710, 227
	
\bibitem[{{Turolla} {et~al.}(2011){Turolla}, {Zane}, {Pons},
{Esposito}, \& {Rea}}]{tzp+11}
  {Turolla}, R., {Zane}, S., {Pons}, J.~A., {Esposito}, P., \&
  {Rea}, N. 2011, ApJ, in press. arXiv:1107.5488
	    
\bibitem[{{Woods} {et~al.}(2001){Woods}, {Kouveliotou},
{G{\" o}{\u g}{\" u}{\c s}}, {Finger}, {Swank}, {Smith}, {Hurley}, \&
 {Thompson}}]{wkg+01}
 {Woods}, P.~M., {Kouveliotou}, C., {G{\" o}{\u g}{\" u}{\c s}}, E., {Finger},
 M.~H., {Swank}, J., {Smith}, D.~A., {Hurley}, K., \&
{Thompson}, C. 2001,  ApJ, 552, 748
		  
\end{thebibliography}
\end{document}